**ORIGINAL ARTICLE**

Antonio Cuéllar · Andrea Ianiro · Stefano Discetti

# Some effects of limited wall-sensor availability on flow estimation with 3D-GANs

**Abstract** In this work we assess the impact of the limited availability of wall-embedded sensors on the full 3D estimation of the flow field in a turbulent channel with $Re_\tau = 200$. The estimation technique is based on a 3D generative adversarial network (3D-GAN). We recently demonstrated that 3D-GANs are capable of estimating fields with good accuracy by employing fully-resolved wall quantities (pressure and streamwise/spanwise wall shear stress on a grid with DNS resolution). However, the practical implementation in an experimental setting is challenging due to the large number of sensors required. In this work, we aim to estimate the flow fields with substantially fewer sensors. The impact of the reduction of the number of sensors on the quality of the flow reconstruction is assessed in terms of accuracy degradation and spectral length-scales involved. It is found that the accuracy degradation is mainly due to the spatial undersampling of scales, rather than the reduction of the number of sensors per se. We explore the performance of the estimator in case only one wall quantity is available. When a large number of sensors is available, pressure measurements provide more accurate flow field estimations. Conversely, the elongated patterns of the streamwise wall shear stress make this quantity the most suitable when only few sensors are available. As a further step towards a real application, the effect of sensor noise is also quantified. It is shown that configurations with fewer sensors are less sensitive to measurement noise.

**Keywords** Turbulent boundary layers · Machine learning · Channel flow · Wall measurements

## 1 Introduction

In this work we challenge the methodology proposed in Ref. [1] for one-shot 3D flow estimation from wall quantities in scenarios with realistic limitations of practical applications. Limited number of sensors, choice of their distribution, measurement noise, and limited physical data information are the main aspects we address in this study.

Many studies have focused on flow control since the work by Prandtl in 1904, where the basics of the physics of boundary layer theory and separation were introduced [2], and great efforts are still being made in this area. Flow control could allow us to design more efficient devices aiming at different purposes, such as maximization of mixing, noise reduction or suppression, heat transfer, lift enhancement, delayed laminar-to-turbulent transition, or skin friction reduction [3].

A. Cuéllar (✉) · A. Ianiro · S. Discetti
Department of Aerospace Engineering, Universidad Carlos III de Madrid, 28911 Leganés, Madrid, Spain
E-mail: acuellar@ing.uc3m.es

A. Ianiro
E-mail: aianiro@ing.uc3m.es

S. Discetti
E-mail: sdiscett@ing.uc3m.es

The control of skin friction in wall-bounded flows is a key element for several energy-intensive industries (e.g. aviation), where the reduction of the viscous drag through turbulence control could lead to important savings. Several surface modifications such as LEBUs (large eddy break-up devices) and riblets have been shown to reduce the skin friction drag [4,5]. However, passive devices cannot be tuned specifically for each operating condition; as such, active flow control is investigated as an alternative to passive flow control [6,7]. Active flow control in a closed-loop arrangement allows to design actuation mechanisms specially designed for the control of a given state of the flow or specific flow features, e.g., coherent structures. With this regard, a more detailed comprehension of the flow allowed to shift the actuating techniques under study from energetically expensive brute-force approaches towards small amplitude forcing [8] to control coherent structures, with significant savings also in terms of mass and size. However, to build a closed-loop flow-control system we need the availability of sufficiently fast and accurate sensing and actuation devices. Hence, it is of utmost importance to research and develop sensing systems for flow control purposes.

The requirement of low intrusiveness sets the need to embed sensors within the wall. Space–time correlation within boundary layers can be exploited for this purpose. The existence of strong correlations among the motions within turbulent boundary layers was first discovered using point measurements with hot wires [9,10]. In particular, linear techniques have been widely used to establish a correlation between the wall measurements and the turbulent flow field. The linear stochastic estimation (LSE) [11–13] searches for a linear transfer function from the wall to the flow fields. Other approaches focus on the existence of coherent structures within the flow and aim at identifying a linear relation between coherent flow features and their wall signature. Since the development of the attached-eddy model by Townsend [14], large eddies attached to the wall are considered to be among the dominant energy-containing motions in wall-bounded turbulent flows. Some preliminary descriptions of the large eddies reported an anisotropic behaviour consistent with the presence of counter-rotating vortex pairs [15]. The presence of these characteristic events with certain coherence and spatio-temporal dependence amidst the randomness of turbulence made it possible to address this problem from a statistical point of view. One of the main tools for the identification of coherent structures in turbulence has been the proper orthogonal decomposition [POD, 16]. Reference [17] first employed POD for the decomposition of wall-bounded turbulence. Obtaining a low-order decomposition of wall-bounded turbulence allows us to have a compact description of the most energetic features of the flow which might have a distinct wall signature. A remarkable example of this approach for flow estimation is represented by the extended POD (EPOD) [18–20] which is found to have a comparable reconstruction accuracy with respect to LSE. Although faithful flow estimations in the vicinity of the wall can be obtained with these linear techniques, they are limited as far as the physics involved contains non-linear events, explaining why some studies coped with this problem by combining linear methodologies with the introduction of non-linearities. When dealing with multiple-time-delay estimation of a turbulent channel flow, faithful linear estimations were established within the viscous layer, while nonlinearities enabled accurate estimations even in the buffer region [21]. With a similar reasoning, Kalman-filter-based estimators extended to account for non-linearities and introduced non-linear forcing were shown to outperform the traditional linear estimators [22,23]. Recently, different resolvent-based approaches have been applied for estimation purposes, including the space-time flow statistics estimation from limited data [24]. The resolvent-based estimation of the non-linear forcing terms improved the estimation obtained with Kalman filter-based estimators [25]. This type of methodology was also used for velocity and pressure estimation in the flow from measurements of pressure and shear stress at the wall [26]. Likewise, resolvent-mode and resolvent-based approaches have been applied to estimate the flow from velocity sensors arranged in given planes of the domain [27].

The capability of machine learning to deal with non-linear problems leaves this branch of methodologies in an outstanding position to describe the non-linear relations needed to estimate the flow within a turbulent boundary layer. Convolutional neural networks (CNNs), thanks to the use of convolutional filters, can detect patterns within image data at multiple scales. Analogously to POD, CNNs can identify patterns within the flow and have been exploited for instance for encoding purposes [28]. The use of convolutional filters within a CNN was shown to have better reconstruction capabilities than EPOD for the task of instantaneous field estimation [20]. While Ref. [20] employed CNNs to estimate the POD coefficients of the flow fields, later works successfully employed fully-convolutional networks (FCNs) [29,30] to reconstruct the flow fields on planes at a certain distance from the wall. These neural networks are capable of making instantaneous reconstructions of the fluctuation velocity field in wall-parallel planes providing them with wall measurements of quantities such as the wall pressure, the wall shear stresses or the heat transfer through the wall. Alternatively, field predictions in other physical configurations have been targeted by CNNs with generative adversarial training, referred to as generative adversarial networks (GANs) [31]. GANs have shown to be very effective also for resolution enhancement purposes [32–34]. The implementation of GANs allowed us to estimate the turbulent field of a channel flow from coarse wall measurements at the same instant [35], reporting an outstanding performance under low-resolution wall-data input and better prediction capabilities than FCNs.

3D convolutional networks have been explored for the purpose of 3D reconstructions of free-surface flows [36] and turbulent channels [33] employing surface and flow field measurements, respectively. We recently extended the instantaneous flow estimation in wall-parallel planes though a GAN [35] to a full 3D estimation with 3D-GANs [1]. This alternative conceptualization of the problem overcomes the need to develop individual planar reconstructions at different wall-normal distances to get a full reconstruction, and produces benefits such as the reduction of trainable parameters in relative terms, or the generation of a 3D field without discontinuities. The availability of 3D fields could be directly used to study coherent turbulent structures. The analysis conducted over the predicted set of Q-events [1] highlighted that the capability of the network to reconstruct the field at a certain location depends both on the distance to the wall and on the wall footprint of the specific structure to reconstruct. Wall-attached sweeps and ejections that penetrate deep towards the centre of the channel can be reasonably well predicted by the 3D-GAN, while detached coherent events might be unseen. The capabilities of the methodology were also compared with traditional linear techniques, such as LSE or EPOD, showing clear advantages over them. The level of error was significantly reduced and the region from the wall to the point with maximum error was clearly extended. The interested reader can refer to Ref. [1] for an extensive performance assessment against linear techniques.

Another concern in flow control studies involves the number and positioning of the sensors. Fewer sensors would be preferred as the computations executed before any actuation or decision may be accelerated, while more sensors might provide a more detailed description of the state of the flow if they are placed efficiently. Even if the flow field to be controlled is a high-dimensional system, coherent structures could be represented on a latent low-dimensional attractor, enabling sparse sensing [37]. Recently developed systems for optimal placement of sensors and actuators have been demonstrated to provide substantially better performances than random placement approaches [38].

The practical implementation of this type of technology for the development of active control strategies might raise questions about its technical complexity or even its feasibility. To get an accuracy comparable to that in Ref. [1], two shear-stress and one pressure sensors must be embedded in the wall within a grid of $64 \times 64$ points (streamwise $\times$ spanwise) arranged within an area equal to $\pi h$ (streamwise) $\times \pi h/2$ (spanwise), with $h$ the half-channel height. Setting this case as a baseline, the objective of this work is to define alternative cases with a reduced amount of sensors and limited physical data information to assess how the performance of the 3D-GAN estimator degrades. Moreover, this work contains an assessment considering noise in measurements, a common issue in instrumental applications with sensors. After this introduction, §2 presents the methodology followed in this work, §3 reports and discusses the results obtained and §4 summarizes this manuscript highlighting the main conclusions and the most remarkable aspects.

## 2 Methodology

This work focuses on the reconstruction of the flow field within a turbulent channel. The dataset includes both wall and flow fields and it is the same dataset employed in Ref. [1], here briefly described in Sect. 2.1. To study the effect of limited sensor number, reduced physical quantity availability, and measurement noise, the information at the wall is subsampled and/or corrupted as described in Sects. 2.1.1 and 2.1.2.

2.1 Turbulent channel flow dataset and sensors in the wall

The dataset has been generated with a DNS [39] of a periodic turbulent channel flow with friction Reynolds number $Re_\tau = 200$, with dimensions $[\pi h \times 2h \times (\pi/2)h]$—respectively in the streamwise, wall-normal, and spanwise directions, indicated respectively with $x$, $y$ and $z$. It contains the three velocity components of the flow field with 64 equispaced points both along the streamwise and spanwise directions, and 64 points in the wall-normal direction from one of the walls to the mid-plane (consider only half of the channel). Moreover, it contains wall measurements of pressure $p_w$ and streamwise and spanwise wall shear stresses $\tau_{w_x}$ and $\tau_{w_z}$. The configuration used in the present work is equivalent to that of Case A of Ref. [1], estimating the flow field of half channel, from the wall to the mid-plane, as represented in Fig. 1. The components of the velocity fluctuations at each wall-normal position are identified as $[u, v, w]$, using the $+$ symbol to indicate quantities expressed in inner scaling.

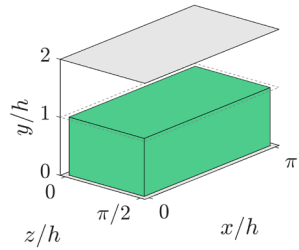

**Fig. 1** Representation of the region from the bottom wall to the mid-plane that is reconstructed providing wall measurements of the bottom wall

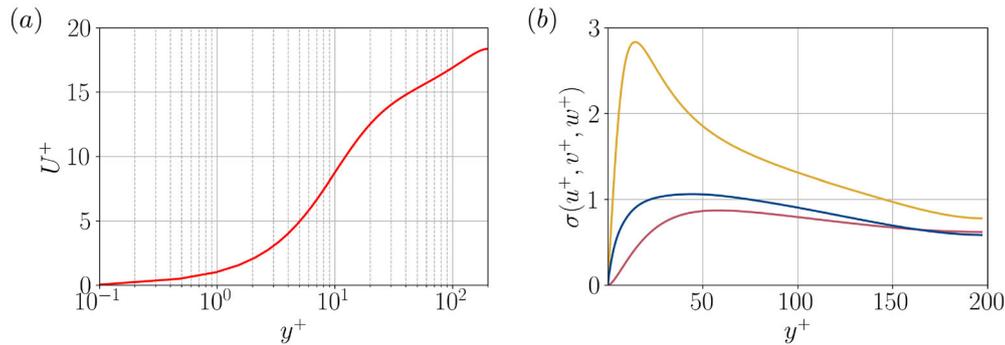

**Fig. 2** Wall-normal profiles of (left) the mean streamwise velocity and (right) standard deviation $\sigma$ of the three velocity components. This figure is adapted from Cuéllar et al., 'Three-dimensional generative adversarial networks for turbulent flow estimation from wall measurements', doi:10.1017/jfm.2024.432, licensed under CC BY 4.0, Ref. [1]

**Table 1** Number of sensors and retained variance with respect to the DNS resolution on each sensor arrangement on the wall

| Sensor arrangement | $64 \times 64$ (%) | $32 \times 32$ (%) | $16 \times 64$ (%) | $16 \times 16$ (%) | $8 \times 32$ (%) | $8 \times 8$ (%) |
|---|---|---|---|---|---|---|
| Number of sensors | $2^{12} \cdot 3$ | $2^{10} \cdot 3$ | $2^{10} \cdot 3$ | $2^{8} \cdot 3$ | $2^{8} \cdot 3$ | $2^{6} \cdot 3$ |
| Retained variance $p_w$ | 100 | 96.6 | 90.0 | 85.3 | 69.8 | 59.7 |
| Retained variance $\tau_{w_x}$ | 100 | 96.9 | 98.3 | 86.9 | 92.0 | 63.1 |
| Retained variance $\tau_{w_z}$ | 100 | 90.0 | 89.9 | 68.9 | 70.1 | 43.2 |

The wall-normal profiles of the channel are represented in Fig. 2. According to the literature, despite being small compared to some other databases, the present channel is big enough to contain self-sustained turbulence [40,41]. Further details comparing this channel with similar channels at $Re_\tau = 180$ can be found in Ref. [1].

*2.1.1 Downsampling procedure*

In this work, we are proposing different grid arrangements for the wall sensors used in the training and estimation process. Table 1 summarizes the different cases proposed in relation to the number of sensors used in each of them and the percentage of retained variance with respect to that of the DNS for each wall quantity. The variance is reduced because the downsampled fields employ bigger sensors which average the measurements of four contiguous sensors in the higher-resolution case, thus filtering smaller turbulent features. The final objective for all cases is to estimate the entire 3D flow field of the half channel, in its full resolution of $64 \times 64 \times 64$ gridpoints. The sensor resolution, on the other hand, is progressively downsampled in powers of two, leading to these three new cases: $32 \times 32$, $16 \times 16$ and $8 \times 8$ (streamwise $\times$ spanwise), as seen in Fig. 3. It represents the patterns' size in relation to the spacing between sensors and illustrates the effect of spatial averaging. Each downsampling step reduces the number of sensors by 75%. The case with full resolution of the wall sensors ($64 \times 64$) is included for reference. It is worth remarking that each test case is trained separately with different random seeds of the weights. This ensures that the effect of sensor number reduction is properly isolated.

The corresponding downsampled sensor inputs are computed with an average pooling filter applied to the original DNS wall data. This is done to maintain full coverage of the wall but with progressively larger (and not overlapped) sensors. This introduces an additional challenge of progressive amplitude modulation of the smallest scales observed in the wall patterns.

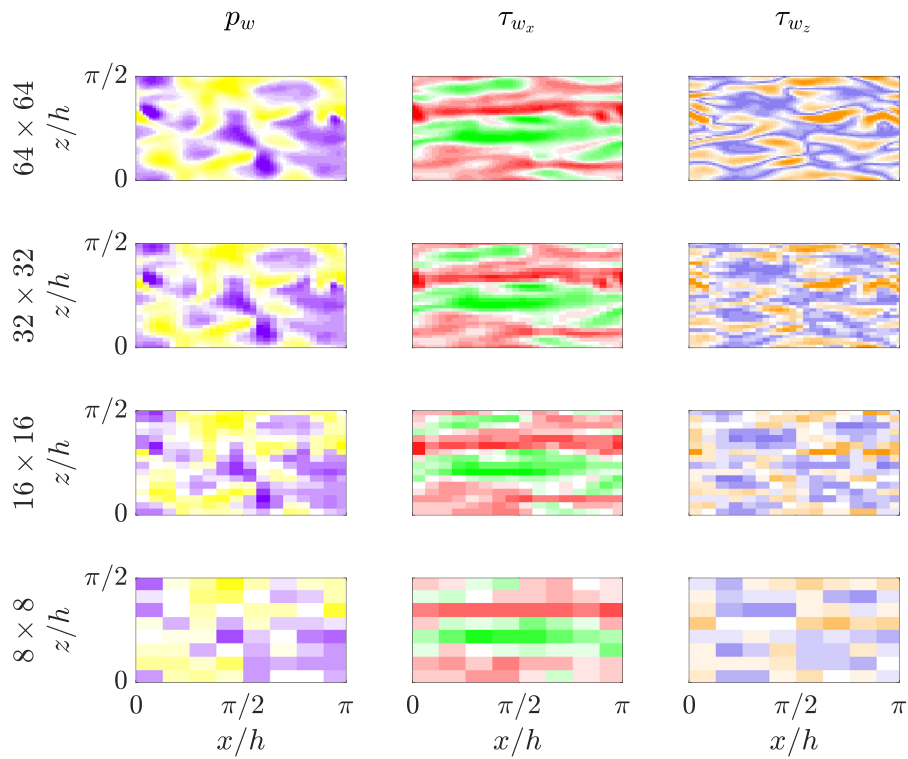

**Fig. 3** Instantaneous representation of wall quantities with different sampling levels. From left to right, wall pressure $p_w$, streamwise and spanwise wall shear stress $\tau_{w_x}$, $\tau_{w_z}$, respectively. From top to bottom, full resolution $64 \times 64$, and progressive isotropic downsampling of factor 2, 4, and 8

Additional cases are proposed in which measurements from wall sensors are not downsampled by the same factor along $x$ and $z$. This is particularly interesting as streamwise-elongated patterns are dominant. An increment in the spanwise sampling resolution could provide further information about the different elongated streaks that are seen along an array of sensors in this direction. On the contrary, increasing the sampling resolution along $x$ might just provide more information about only one or a few streaks. Smart sensor placement techniques have been shown to use the same amount of sensors more efficiently [37,38]. The methodology proposed here is not a smart sensor placement but tries to compare homogeneous sensor patterns with the same number of sensors providing full coverage of the domain considered in this problem. In line with this asymmetric downsampling, two additional cases are proposed, with corresponding baseline with the same number of sensors but symmetric distribution: $16 \times 64$ (being $32 \times 32$ as baseline) and $8 \times 32$ (with $16 \times 16$ as a reference for comparison). A visual representation of this downsampling scheme can be seen in Fig. 4.

To consider the case of limited physical data information we also explore the effect of employing sensors capable of measuring only one physical quantity—only $\tau_{w_x}$, $\tau_{w_z}$ or $p_w$. Furthermore, we address jointly the effect of the lack of availability of high sensor resolution in the wall, together with that of different quantities.

### 2.1.2 Noise modelling

In real applications, measurement noise in the sensors must be considered. To simulate this effect affecting wall measurements, random Gaussian noise is added to $p_w$, $\tau_{w_x}$ and $\tau_{w_z}$. Two levels of noise are tested on the four cases with the same number of sensors in both directions. The standard deviation of the noise $\delta$ is set to 1% and 3% of the standard deviation of each of the wall-measured quantities. Recall that the downsampling is applied in terms of an average pooling layer directly applied to the sensed data. For these cases, the level of noise is thus increased according to $\sqrt{N}$ (see Table 2) to account for the smoothing effect of the average pooling. No bias error is considered for simplicity.

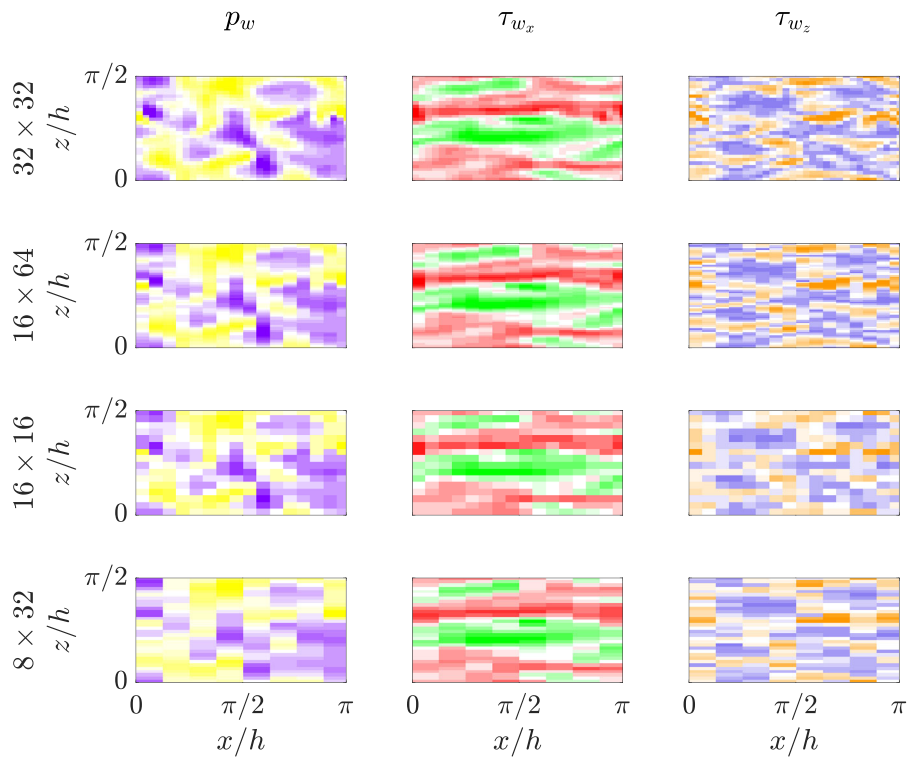

**Fig. 4** Instantaneous representation of wall quantities with different sampling levels. From left to right, wall pressure $p_w$, streamwise and spanwise wall shear stress $\tau_{w_x}$, $\tau_{w_z}$, respectively. From top to bottom, on even lines the cases with asymmetric sensor distribution, and on odd lines their corresponding baseline cases

**Table 2** Number of sensors used per downsampled cell

| Sensor arrangement | 64 × 64 | 32 × 32 | 16 × 16 | 8 × 8 |
|---|---|---|---|---|
| N | 1 | 4 | 16 | 64 |

To estimate the effect of noise on the reconstruction quality, isolating training and reconstruction processes, the networks, previously trained with the clean DNS data, are fed with the wall fields with added noise. Results are reported in Sect. §3.4 where the performance is also compared with that of the reconstruction without noise in the input data.

2.2 Generative adversarial network

The flow estimation is carried out with a GAN. The architecture comprises two networks, the generator $\mathcal{G}$ and the discriminator $\mathcal{D}$. The generator is fed with 2D instantaneous sensor fields and provides the three 3D fields of $u$, $v$ and $w$ at the same time instant. The discriminator is used solely during the training process. It is fed with flow fields—either the original ones from the DNS or those generated by $\mathcal{G}$—and must classify them accordingly. The two networks are trained adversarially. The loss function introduces a penalty based on the adversarial loss, quantifying how good is the performance of each network in the "game" it is playing against the other. At the end of the training process, $\mathcal{G}$ should be capable of generating turbulent flow fields with a level of accuracy such that $\mathcal{D}$ is confused, having difficulty in classifying fields as original or generated.

The GAN architecture we used is based on the field estimator in wall-parallel planes developed in Ref. [35]. It was extended to 3D in Ref. [1], with 64 × 64 sensor-grid inputs onto $\mathcal{G}$. In this work, the architecture is adapted to accommodate different input sizes (see sketch in Fig. 5). Upsampling layers are applied to make the size of the output of $\mathcal{G}$ match the expected 64 × 64 × 64 resolution of the half-channel field. Along the $y$ direction, the domain needs to grow from a unique wall-parallel layer to 64, for which six upsampling layers of factor 2 are employed [1]. Instead, along both $x$ and $z$, the sizes of the domain need to be widened by a factor defined as the ratio of the output (64) and the corresponding input sizes (64 − 32 − 16 − 8). This step is done with a single layer introduced after the first convolution filter of the 3D-GAN network in [1]. Alternatively, an additional procedure to increase the resolution of the wall measurements before feeding them to the 3D-GAN could be considered. However, CNNs can be used with super-resolution purposes [42], and the architecture employed in $\mathcal{G}$ may jointly allow to increase the resolution and predict the flow field from wall measurements at once. The use of the same architecture for the different test cases, without the need of additional networks or any other type of implementation, and having used the same amount of trainable parameters, further allows us to establish a fairer comparison between them.

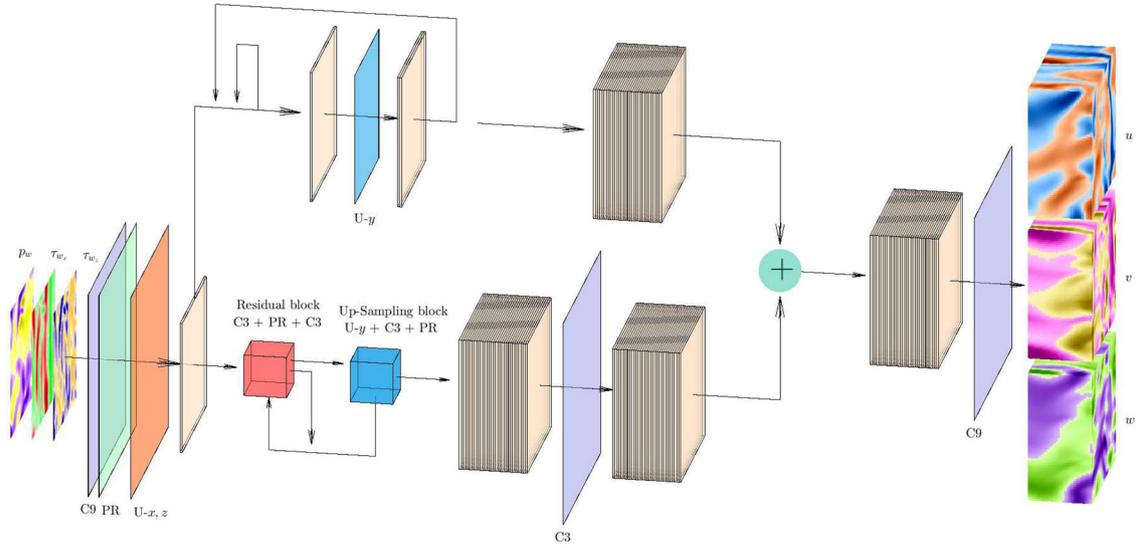

**Fig. 5** Diagram of the generator network. The network contains convolutional layers (C3 and C9 with kernel size 3 or 9 respectively), parametric-ReLU (PR) and upsampling layers acting on the wall-normal direction (U-y) and in both wall parallel directions (U-x,z). The residual block and the up-sampling block are recursively repeated. This figure is adapted from Cuéllar et al., 'Three-dimensional generative adversarial networks for turbulent flow estimation from wall measurements', doi:10.1017/jfm.2024.432, licensed under CC BY 4.0 (https://creativecommons.org/licenses/by/4.0/), Ref. [1]

Moreover, the network is a deep neural network consisting of convolutional layers and PReLU activation functions [43]. They are arranged along 32 residual blocks with skip connections. The six upsampling layers to widen the domain in the $y$ direction are spread along the depth of the network, and are followed by an additional convolutional filter. After all the residual blocks, there is a global skip connection and a final convolution layer with three output filters representing the [$u$, $v$, $w$] velocity fluctuations respectively.

The discriminator receives the fields of [$u$, $v$, $w$] with resolution $64 \times 64 \times 64$, and the size of the domain is progressively reduced, first with a set of convolutional filters with striding, and then with two fully-connected layers, with 1024 and 1 neuron as output. Finally, a sigmoid activation function is used to classify the field providing an output in the $[0 - 1]$ range. A sketch of the discriminator network $\mathcal{D}$ is presented in figure 4 of Ref. [1].

The loss function of each network is composed of two terms. One aims to improve its own performance in terms of fitting the training data, while the other introduces a penalty based on the performance of the other network (the so-called adversarial loss). The loss function of $\mathcal{G}$ contains the MSE of the generated velocity field with respect to the original field from the DNS ($\mathcal{L}_{\text{MSE}}$), and the binary cross-entropy as adversarial loss:

$$\mathcal{L}_G = \mathcal{L}_{\text{MSE}} - 10^{-3}\mathbb{E}[\log D(\boldsymbol{u}_{\text{GAN}})], \quad (1)$$

where $\mathbb{E}$ is the expectation operator and $D(\boldsymbol{u}_{\text{GAN}})$ represents the output of $\mathcal{D}$ when it receives a GAN-generated velocity field.

The loss function for $\mathcal{D}$ is defined as the sum of two terms: its own expectation when it receives a DNS field as content loss and the expectation when it receives a generated field as adversarial loss:

$$\mathcal{L}_D = -\mathbb{E}[\log D(\boldsymbol{u}_{\text{DNS}})] - \mathbb{E}[\log(1 - D(\boldsymbol{u}_{\text{GAN}}))]. \quad (2)$$

**Table 3** Integral error $\varepsilon$ for the different cases proposed

| Grid | 3 sensor types | 1 sensor type | | | 3 sensors + noise | |
| --- | --- | --- | --- | --- | --- | --- |
| | | $p_w$ | $\tau_{w_x}$ | $\tau_{w_z}$ | 1% | 3% |
| $64 \times 64$ | 0.207 | 0.224 | 0.232 | 0.242 | 0.219 | 0.234 |
| $32 \times 32$ | 0.211 | 0.233 | 0.241 | 0.254 | 0.222 | 0.239 |
| $16 \times 64$ | 0.214 | – | – | – | – | – |
| $16 \times 16$ | 0.227 | 0.272 | 0.2664 | 0.321 | 0.228 | 0.232 |
| $8 \times 32$ | 0.229 | – | – | – | – | – |
| $8 \times 8$ | 0.262 | 0.318 | 0.293 | 0.357 | 0.262 | 0.262 |

During training, all DNS nodes are assigned the same weight in the loss, regardless of their corresponding volume (the mesh is finer close to the wall). We observed in our previous work [1] that the accuracy of the flow reconstruction is mainly driven by the existence of certain relations between wall and flow data. Consequently, introducing weights based on the volume of each DNS cell does not have a significant impact on training, and might even result in a detrimental effect on the accuracy, especially in the near-wall region.

## 3 Results

The mean-squared error (MSE) is used to evaluate the performance of $\mathcal{G}$ in estimating the fluctuation velocity fields. It is computed independently for each component $[u, v, w]$ and for each $y^+$ coordinate and discussed in the following for the cases under study according to:

$$MSE(u_i^+, y^+) = \frac{1}{N_s N_x N_z} \Sigma_s \Sigma_x \Sigma_z (u_{i,\text{DNS}}^+(i,j,k) - u_{i,\text{GAN}}^+(i,j,k))^2 \qquad (3)$$

where $u_i^+$ represents each possible velocity component $[u^+, v^+, w^+]$ with the subscript DNS or GAN to represent respectively the original and the reconstructed velocity fields, $N_s$ is the number of samples in the testing set and $N_x$ and $N_z$ are the number of grid points along $x$ and $z$, respectively.

Additionally, an integral error $\varepsilon$ metric (4) based on the MSE has been defined to facilitate a quick comparison between the performance of different cases, merging in a single number the effect of each component and the distance to the wall. Table 3 collects the metric for the different cases considered in this work:

$$\varepsilon = \frac{1}{200} \int_{y^+=0}^{y^+=200} \frac{MSE(u) + MSE(v) + MSE(w)}{[rms(u) + rms(v) + rms(w)]^2} dy^+ . \qquad (4)$$

Results are also discussed with a spectral analysis of the wall-sensed scales and the velocity scales.

3.1 Estimation with same downsampling factor in $x$ and $z$

The performance of the four cases with the same downsampling factor in both directions can be compared in Fig. 6.

First of all, it should be remarked that all the curves observed in Fig. 6 have a similar behaviour. The error is very low in the vicinity of the wall, it reaches a maximum and then decreases. Practically no differences in error are observed at the centre of the channel. As the sensor resolution becomes coarser, the position of the maximum error shifts towards the wall, especially for $u^+$. For all these cases, the MSE of $u^+$ is the highest, with maximum values of about $u_\tau$ and even $2u_\tau$ for $8 \times 8$. The MSE of $w^+$ is in an intermediate position, with the MSE of $v^+$ being the lowest. This is in line with the magnitudes of the standard deviation of the velocity fluctuations, as reported in Ref. [1].

As expected, the error increases progressively as the wall measurements become coarser—following cases with the same downsampling factor along both $x$ and $z$ directions. However, the additional error introduced between cases is not the same. From a qualitative observation of Fig. 3, one could notice that the smallest patterns observed with the $64 \times 64$ resolution are nearly preserved for the downsampled set with $32 \times 32$ sensors. Significant scale losses are instead observed for the $16 \times 16$, and in particular, for the $8 \times 8$ resolutions. Figure 6 confirms that the $32 \times 32$ wall input generates a slightly higher error than that with $64 \times 64$ sensors, while these jumps increase when the resolution is further reduced. These jumps are in line with the behaviour of the integral error $\varepsilon$ reported in Table 3. This effect should be assessed in relation with the distance between sensors (see Table 4) and the Power Spectral Density (PSD) of the measured patterns at the wall (Fig. 7). On the one hand, regarding the spacing between the sensors, the minimum characteristic sizes of the streamwise and spanwise patterns expected to be retained are twice the spacing distances reported in Table 4, according to the Nyquist-Shannon sampling theorem. Except for some of the smallest and low-energetic patterns observed in the 64 × 64 maps in Fig. 3, the sensor spacing for the 32 × 32 arrangement (see Table 4) fulfils this condition for most of the relevant flow scales. This might explain why the loss in accuracy reported in Fig. 6 between the 64 × 64 and 32 × 32 arrangements is minimum—despite the strong reduction in the number of sensors from 12288 to 3072. Instead, the 16 × 16 sensor spacing leaves an important proportion of the patterns out of the threshold; this effect is further exacerbated for 8 × 8 sensors. Similarly, Table 1 shows how the retained variance with respect to the original resolution, as an indicator of the contained energy, is quite similar for the cases with 64 × 64 and 32 × 32 sensors, while it is reduced more noticeably for 16 × 16 and 8 × 8 sensors. These reductions in the number of sensors determine an inadequate sampling of relevant flow scales, losing patterns that seem to be important for the reconstruction of the 3D flow field, in particular for the $u$ component.

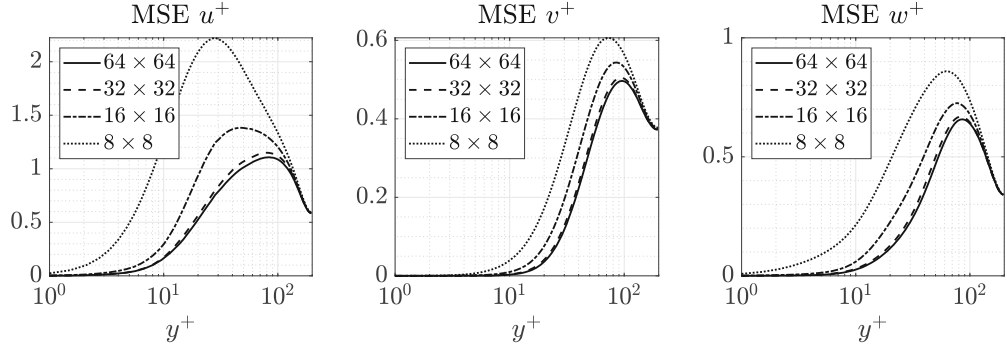

**Fig. 6** MSE reported for each velocity component and input-resolution case as a function of the wall-normal distance (———, 64 × 64; - - -, 32 × 32; –·–·, 16 × 16; ······, 8 × 8). The results are expressed in wall-inner units based on the friction velocity $u_\tau$

**Table 4** Spacing between gridpoints in the wall of the channel for each resolution configuration employed

|  | 64 | 32 | 16 | 8 |
| --- | --- | --- | --- | --- |
| $\Delta x/h$ | 0.049 | 0.098 | 0.196 | 0.393 |
| $\Delta x^+$ | 9.82 | 19.63 | 39.27 | 78.54 |
| $\Delta z/h$ | 0.025 | 0.049 | 0.098 | 0.196 |
| $\Delta z^+$ | 4.91 | 9.82 | 19.63 | 39.27 |

On the other hand, the spectrum maps shown in Fig. 7 (top row) also denote that large patterns—both along $x$ and $z$—dominate for the three measured quantities, and those of $\tau_{w_x}$ are particularly elongated in the streamwise direction. The $p_w$ and $\tau_{w_x}$ spectra are quite well conserved, with some degree of distortion growing with the downsampling factor, but still moderate for 16 × 16. The fact that both the 8 × 8 $\lambda_x^+$ and $\lambda_z^+$ minimum wavelenghts penetrate within the contoured region (10% threshold) of the 64 × 64 map truly constraints the original spectral distribution for this case. The preservation of the $\tau_{w_z}$ maps is worse, mainly because the distribution of its characteristic scales reveals shorter and thinner patterns than for the other quantities. The downsampling has an effect on the minimum wavelength sampled. For what concerns the case with 32 × 32 resolution the effect is mainly felt in the streamwise direction. For the 16 × 16 resolution, the minimum wavelengths penetrate within the contoured region both in the streamwise and the spanwise direction. This effect is of course stronger for the 8 × 8 resolution. Beyond this limitation, the contoured lines within their respective limits look distorted to a greater extent than for the other two quantities. This discussion suggests that the penalty of these downsampled sensor arrangements affects the field reconstructions more significantly with regard to the role of $\tau_{w_z}$.

Moreover, the spectral length-scales of the streamwise velocity component that the 3D-GAN is able to reconstruct from the different input sensor arrangements are reported in Fig. 8. In the different wall-parallel sections of the channel analyzed, the dominant patterns present in the original DNS field are as elongated as $\pi h$, with a width of about 100–150 wall inner units. At a distance of 15 wall units away from the wall,

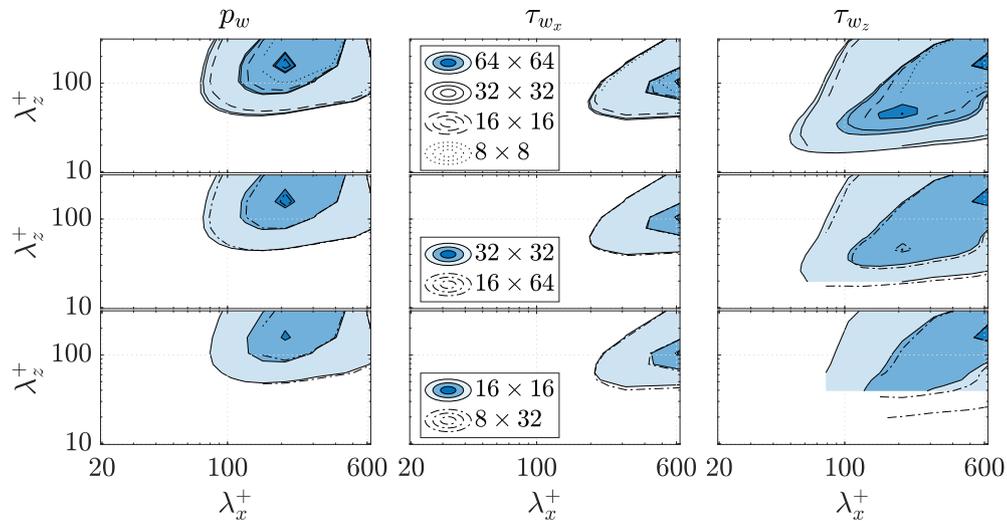

**Fig. 7** Premultiplied PSD maps of the wall-sensed $p_w$ (left), $\tau_{w_x}$ (centre) and $\tau_{w_z}$ (right). Cases with different sensor arrangements (top row: coloured ——, $64 \times 64$; ——, $32 \times 32$; - - -, $16 \times 16$; ······, $8 \times 8$; central and bottom rows: coloured ——, $64 \times 64$ and $32 \times 32$; –·–·, $16 \times 64$ and $8 \times 32$) are compared along each row according to the legend. The contour levels correspond to 10%, 50% and 90% of the maximum PSD level in the original $64 \times 64$ map

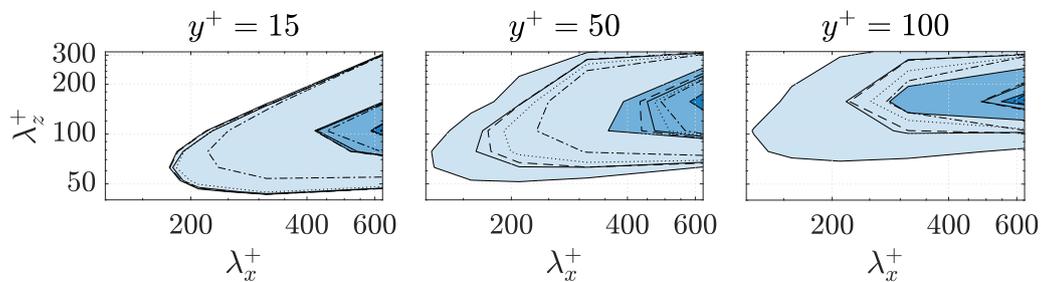

**Fig. 8** Premultiplied PSD maps of the streamwise velocity component $u^+$ at three different wall-normal distances. The estimations from different number of sensors (contoured lines: ——, $64 \times 64$; - - -, $32 \times 32$; ······, $16 \times 16$; –·–·, $8 \times 8$) are compared with the spectrum of the DNS (coloured contour: ——). The contour levels correspond to 10%, 50% and 90% of the maximum PSD level in the DNS $u^+$ field

the spectra look quite well preserved, with a certain attenuation for the $8 \times 8$ arrangement. The spectrum for $16 \times 16$ is nearly coincident with both finer resolutions, even if its MSE ($u^+$) at this distance from the wall is substantially larger (see Fig. 6). At $y^+ = 50$ the spectral maps have a significant additional distortion, with an important part of the spectral power not recovered: no point in the map gets the 90% threshold even at the $64 \times 64$ resolution; this effect is more intense for coarser wall data. The attenuation is such that the $64 \times 64$ map loses almost 50% of the spectral power contained in the DNS data. The attenuation observed at $y^+ = 100$ is even strengthened, leading to a loss of more than 70%. A more detailed quantification of this attenuation is reported in Table 5. Being $S$ the premultiplied power spectral density map of the predicted field, and $\hat{S}$ that of the original field, the overall attenuation is $1 - \frac{\Sigma(S)}{\Sigma(\hat{S})}$, and the highest spectral intensity attenuation is $1 - \frac{max(S)}{max(\hat{S})}$. These results show the robustness of the 3D-GAN when using significantly fewer sensors than in the baseline configuration. The attenuation at $8 \times 8$ with respect to the full resolution is evident, but the reconstruction is certainly satisfactory if one considers that it employs only 1.5% the number of sensors in the $64 \times 64$ arrangement. It must be remarked, however, that for regions further from the wall the reconstruction error is mainly driven by the lower correlation of velocity fluctuations with wall quantities, thus the effect of reducing the number of sensors is much less relevant.

Additionally, the turbulent structures have been analyzed. Individual Q-events have been identified with the 3D-quadrant analysis [44] based on the planar quadrant analysis for the identification of turbulent structures

**Table 5** Attenuation of the $u^+$ field spectrum with respect to the DNS at $y^+ = [15, 50, 100]$

| Overall attenuation | | | | | Highest spectral intensity attenuation | | | |
|---|---|---|---|---|---|---|---|---|
| $64 \times 64$(%) | $32 \times 32$ (%) | $16 \times 16$ (%) | $8 \times 8$ (%) | | $64 \times 64$ (%) | $32 \times 32$ (%) | $16 \times 16$ (%) | $8 \times 8$ (%) |
| 5.8 | 4.3 | 11.3 | 39.6 | $y^+ = 15$ | 0.0 | 0.0 | 0.3 | 8.2 |
| 46.0 | 47.0 | 57.9 | 70.3 | $y^+ = 50$ | 18.3 | 11.8 | 22.3 | 24.7 |
| 71.5 | 72.2 | 81.7 | 83.0 | $y^+ = 100$ | 33.4 | 30.7 | 54.4 | 44.9 |

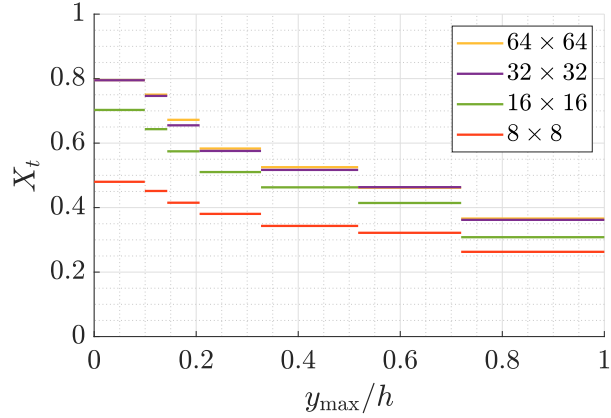

**Fig. 9** Evolution of the overall matching proportion $X_t$ of the wall-attached structures in the DNS fields with those from the generated fields. Structures are categorized into different groups according to their span $y_{max}/h$

[45,46]. Structures are defined according to a hyperbolic hole size equal to H=1.75 following:

$$| - u(x, y, z)v(x, y, z)| > Hu'(y)v'(y) , \qquad (5)$$

where $u'(y)$ and $v'(y)$ represent the root-mean-squared of each velocity component at a given wall distance $y$. These events have been bounded by their range in the wall-normal direction, being $y_{min}$ and $y_{max}$ the minimum and maximum distance from the wall of the points forming a structure, respectively. Wall-attached structures (setting the threshold at $y_{min}^+ \leq 20$) represent the vast majority of the population of turbulent structures. For each wall-attached structure in the DNS dataset, the proportion of its volume $\mathcal{V}$ that is overlapped with analogous Q-events from the corresponding estimated (3D-GAN) field has been computed.

Structures have been classified in bins according to their $y_{max}$. For each bin, the average volume matching proportion $X_t$ of all its structures has been computed as the average of the average overlap proportion $\mathcal{V}$ of all these structures [1]. Figure 9 shows two main effects regarding $X_t$. Overall, wall-attached structures are preserved better the closer to the wall they remain, while those that extend towards the center of the channel have more chances to lose a bigger part of them in the reconstruction process. Besides, it shows the trend due to the effect of the number of sensors used. In line with some previous comments, the quality of the estimation becomes worse as fewer sensors are employed for the whole $y_{max}/h$ span. However, the trend is not uniform, with the configuration with $32 \times 32$ sensors suffering a small penalty with respect to $64 \times 64$, while Q-events generated with cases with further downsampling report more significant losses in terms of $X_t$. This is probably due to the aforementioned effect of the downsampling on the wall spectra and the length-scales that each sensor set is able to measure.

3.2 Estimation with different downsampling factor in $x$ and $z$

Alternative cases were proposed applying different downsampling factors along directions $x$ and $z$ (see Fig. 4). The MSE of the two new cases can be observed in Fig. 10, where they are compared with their respective cases with the same number of sensors with equal downsampling factors in $x$ and $z$. As seen in Table 1, the $16 \times 64$ arrangement has the same number of sensors as the $32 \times 32$ arrangement, while the same holds for the cases with $8 \times 32$ and $16 \times 16$ sensors.

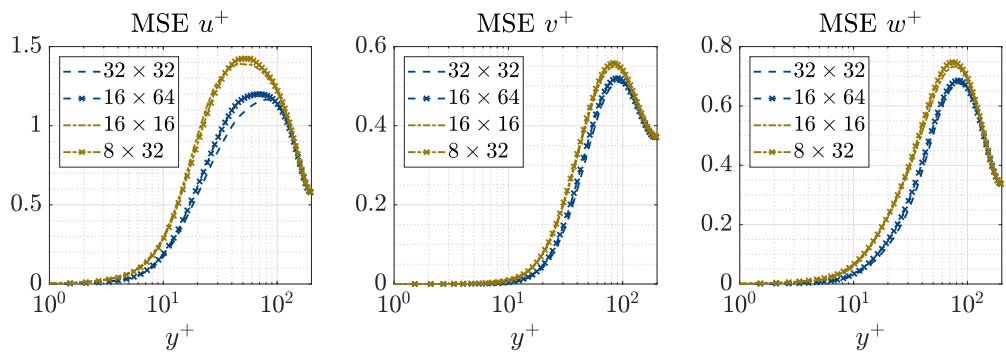

**Fig. 10** MSE comparison for each velocity component as a function of the wall-normal distance for cases with the same number of sensors (- - -, 32 × 32; ×- - -, 16 × 64; –·–··, 16 × 16; ×-·–··, 8 × 32). The results are expressed in wall-inner units based on the friction velocity $u_\tau$

The difference in error between these two pairs of cases is quite small. It is seen how the sensor arrangements with the same downsampling factor in $x$ and $z$ indeed work slightly better than those with different downsampling factors, against the hypothesis introduced in §2—except for the estimation of $u^+$ with $2^{10} \cdot 3$ sensors in the range of $y^+ \approx [15 - 30]$. As in the DNS grid, all cases with the same downsampling factor already have twice as much resolution in $z$ as in $x$. On the other hand, in the alternative arrangements, the resolution in $z$ is 8 times as in $x$. There might be a ratio between the resolution along both directions for which the error is minimized. Although a ratio of 8 might seem a bit extreme for this optimum scenario, the penalty in terms of error is not very significant. Besides, given the quantization of this additional penalty as reported in Fig. 10, under the scenario of a physical implementation of this technology, these alternative arrangements could even be considered if they were preferred for any practical purpose against arrangements with same downsampling factor, at the cost of a slightly poorer estimation performance. Similar conclusions could be drawn from the integral error $\varepsilon$ (see Table 3), where both cases report a very moderate increment with respect to the cases with the same number of sensors.

The spectrum of the three wall measurement quantities for these two pairs of cases can be compared in the central and bottom rows of Fig. 7. In both $p_w$ spectra, the distortion of the contoured lines of the alternative arrangements with respect to those with same downsampling factor is not very significant. However, the downsampling factor of 8 applied on $x$ in 8 × 32 highlights spectral losses of streamwise scales—as previously reported in the 8 × 8 arrangement. Regarding the $\tau_{w_x}$, quantity for which scales are particularly elongated, none of these arrangements seems to be significantly penalized and the contoured lines are quite overlapped. Instead, the preservation of the scales in the $\tau_{w_z}$ map is not as good, as this spectrum contains smaller scales that might not be bounded by the minimum wavelengths. The increment in resolution in $z$ in these two alternative arrangements with respect to those with same downsampling factor enables a reduction of the loss of the small $\tau_{w_z}$ spanwise scales previously reported in Sect. 3.1. Indeed, in the 16 × 64 case, the increment in the spanwise resolution with respect to 32 × 32, allows us to partially recover a peak in the spectrum around $\lambda_x^+ \approx 200$ and $\lambda_z^+ \approx 50$ that was only reported in the original 64 × 64 spectrum map. Nevertheless, this is possible at the cost of losing further small streamwise patterns. In relation to the retained variance reported in Table 1, it is seen that the retained variance of the patterns of $\tau_{w_x}$, particularly elongated, is better conserved with 16 × 64 and 8 × 32 than on their respectives cases 32 × 32 and 16 × 16. However, this difference is much smaller than the loss reported for the retained variance of $p_w$.

### 3.3 Reconstruction with one type of sensor

Often, in realistic scenarios only one of the wall quantities might be accessible. In this section, we analyse the effect of reducing the availability of the wall-measured quantities. Figure 11 shows a comparison of the MSE reported when the 3D-GAN is fed only with pressure, streamwise or spanwise wall shear stress, arranged in the 64 × 64 resolution. The error of the baseline case with the three types of sensors is shown here for reference.

These results show that the relative importance of each of these three wall quantities for the flow field reconstruction is not the same. Besides, their relation with each velocity component is different. Close to the wall, the streamwise and spanwise velocity components show a stronger relation with the streamwise and spanwise wall shear stresses, respectively. However, the MSE corresponding to the wall pressure becomes lower beyond $y^+ \approx 25$ for $u^+$ and beyond $y^+ \approx 10$ for $w^+$. Instead, $p_w$ shows a slightly better capability to estimate $v^+$ throughout the domain.

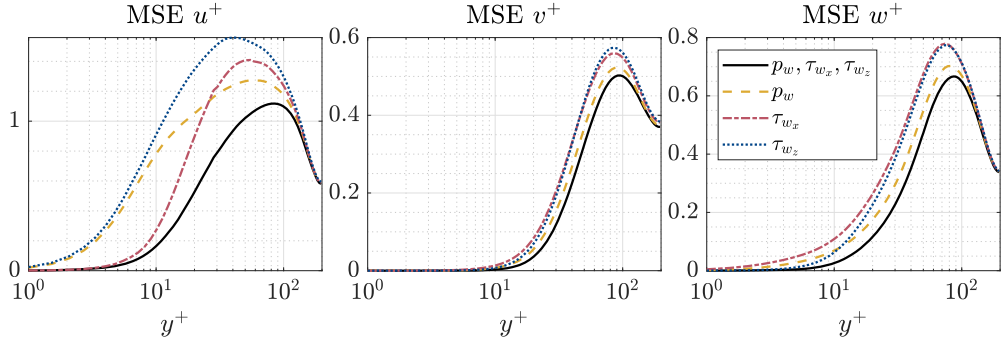

**Fig. 11** MSE comparison of each velocity component between cases using only one type of sensor with the $64 \times 64$ resolution: ——, $p_w, \tau_{w_x}, \tau_{w_z}$ (baseline); - - -, $p_w$; — · —, $\tau_{w_x}$; ······, $\tau_{w_z}$

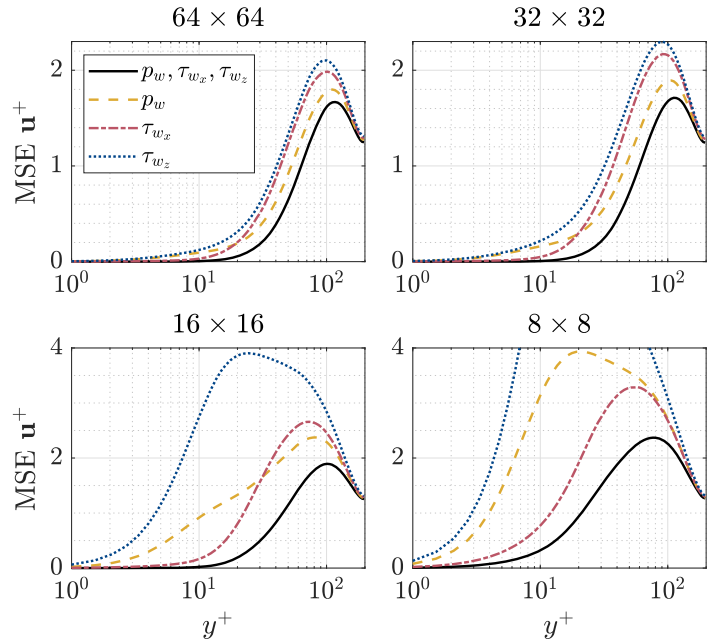

**Fig. 12** MSE comparison between cases using only one type of sensor with different resolutions: ——, $p_w, \tau_{w_x}, \tau_{w_z}$ (baseline); - - -, $p_w$; — · —, $\tau_{w_x}$; ······, $\tau_{w_z}$

A similar procedure is followed for the downsampled resolutions proposed in Fig. 3. To simplify it, we compare the MSE of the fluctuation velocity $\mathbf{u}^+$, defined as the sum of the MSE of $u^+$, $v^+$, and $w^+$, instead of for each of its components separately. The results for the four cases are plotted in Fig. 12.

For the $64 \times 64$ resolution, it is observed that the pressure sensors offer the best estimation performance throughout most of the domain. Only in the vicinity of the wall, up to $y^+ \approx 15$, $\tau_{w_x}$ provides a slightly smaller MSE. This finding is similar to that for the MSE of $u^+$ shown in Fig. 11. As the MSE of $u^+$ for the case using only $\tau_{w_x}$ reports a clear advantage close to the wall against the other two quantities, its influence prevails ($64 \times 64$ in Fig. 12) when the MSE of the other velocity components are included (cases in which $\tau_{w_x}$ does not provide the lowest MSE but the differences between cases employing different wall-measured quantities are less significant).

No significant changes are reported for the $32 \times 32$ resolution, apart from a slight increase in the magnitude of the MSE. This effect was also shown in Fig. 6, where the three wall quantities were employed. However, differences are found for the cases with the lowest number of sensors. Having a limited number of sensors appear to be particularly detrimental for estimation using solely pressure. The estimation based on $\tau_{w_x}$, on the other hand, is still reasonably good in the near-wall region. With $16 \times 16$ sensors, the pressure is not able to provide a lower MSE than $\tau_{w_x}$ until beyond $y^+ = 30$. With $8 \times 8$ sensors, $\tau_{w_x}$ provides the lower MSE throughout the way along from the wall to the mid-plane of the channel. As observed qualitatively in Fig. 3 and quantitatively in Fig. 7, $\tau_{w_x}$ exhibits large-scale patterns, which might explain why this quantity can provide us with better field estimations under a coarse grid arrangement when only one type of sensor is used.

The MSE of $\tau_{w_z}$ for these two coarser resolutions has very significant increments with respect to the two finer resolutions, even if with $64 \times 64$ sensors it already provided the worst estimation. This might be due to the fact that the patterns in $\tau_{w_z}$ are smaller, while the downsampling procedure preserves better the patterns of the other two quantities.

3.4 Effect of sensor noise

This section covers an analysis consisting on the quantification of the flow field estimation MSE when noise is applied over the three wall-sensed quantities.

Figure 13 shows the MSE profiles of the three velocity components (columns) for each resolution considered (rows), comparing this metric for the values of $\delta = [0\%, 1\%, 3\%]$. As expected, the MSE of the reconstructed fields increases with the noise added to the sensors. The curves follow a similar profile, with very low values in the vicinity of the wall, progressively increasing with the distance from the wall up to a maximum, and then reduced and stabilized towards the centre of the channel. For the three values of $\delta$ considered, the MSE levels are very similar in the closest and in the farthest regions to the wall, while the differences in the region in between these two are more evident. All this discussion applies to the three velocity components, although the differences in MSE for $u^+$ are more significant.

In relation with the input resolution, the main conclusion is the fact that the effect of $\delta$ is minimized with the downsampling factor. For the sensor arrangement $32 \times 32$ few differences are reported with respect to $64 \times 64$; the shifts between the curves of each $\delta$ are still very significant, and the level of error with $32 \times 32$ sensors is slightly higher than for $64 \times 64$. With further downsampling, $16 \times 16$ sensors shows more moderate increments of the MSE curves with increasing $\delta$. This effect is even strengthened at $8 \times 8$, where the curves of each level of $\delta$ are nearly overlapped throughout the domain of the channel flow. In other words, all this discussion implies that, as the sensor arrangement becomes coarser, the MSE grows faster for $\delta = 0\%$ than for $\delta = 1\%$, and even more than for $\delta = 3\%$. As a remarkable example, comparing curves with $\delta = 3\%$ (more evident for $u^+$), the MSE for $16 \times 16$ is higher than for $32 \times 32$ and $64 \times 64$ only close to the wall and close to the centre of the channel, but in the rest of the channel and in particular, in the region of the peak, the MSE is significantly lower for $16 \times 16$ sensors. The same trends with the sensor resolution are reported in Table 3 according to the integral error $\varepsilon$.

4 Conclusion

The 3D-GAN developed in Ref. [1] has been tested under coarser grids of input sensors at the wall, willing to estimate the flow of half channel at the same resolution as in the original grid. This is a challenge for the methodology, allowing us to assess its performance when less data are available. This analysis quantifies the additional uncertainty introduced when fewer sensors are installed, which would definitely be a trade-off between simplicity and accuracy to analyze under a practical implementation scenario. The robustness of the 3D-GAN under these scenarios is demonstrated by the fact that the penalization reported under coarser wall measurements is not directly proportional to the reduction in the number of sensors employed, but rather depends on their capability to represent wall patterns.

When a progressive downsampling to the sensor grid is applied in both streamwise and spanwise directions in powers of 2, the three velocity components experience an increasing MSE metric over the whole half channel. Nevertheless, the error introduced with each downsampling step is not linearly dependent on the number of sensors. The quality of the reconstruction seems to be directly affected by the characteristic length-scales of the measured quantities, and which of them remain present with each of the sensor arrangements. The MSE for a wall input of $32 \times 32$ is almost equivalent to that of $64 \times 64$ sensors—the DNS resolution. Nevertheless, the smaller scales lost with any further downsampling seem to affect the accuracy of the estimator. The largest

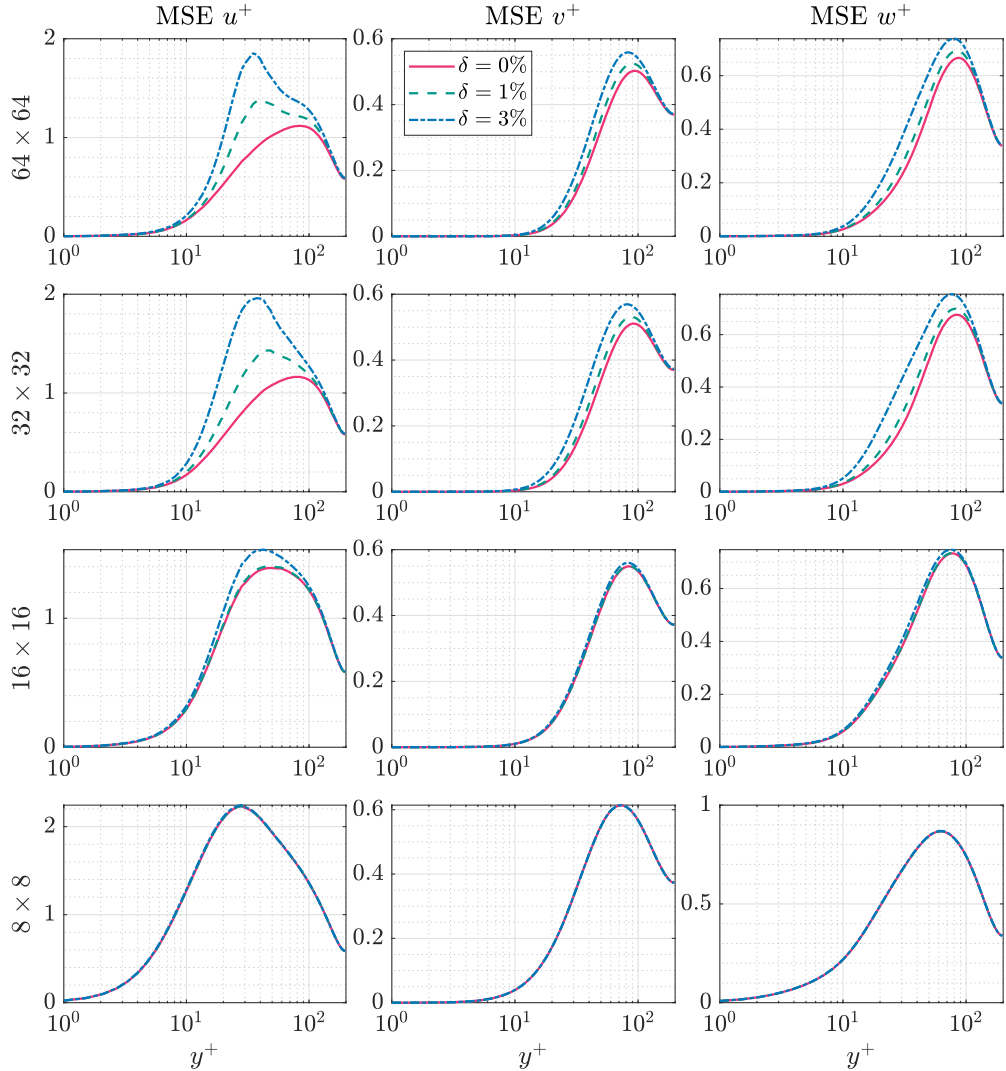

**Fig. 13** MSE comparison between cases with different noise levels in the wall input ——, $\delta = 0\%$; - - -, $\delta = 1\%$; —·—·, $\delta = 3\%$; for each velocity component and resolution case

scales that remain in the wall after the coarsening process still carry very important information for the flow reconstruction. Although the attenuation experienced in the estimated fields becomes relevant, the largest scales present in the channel flow are still sensed.

The aforementioned cases are compared as they are downsampled in the same way along $x$ and $z$. A comparison between this sensor arrangement setting and a different one with different downsampling in the two directions is established. The main hypothesis was that the structures are mostly elongated in the streamwise direction, thus it is more relevant to sample them properly in the spanwise direction. The results show, however, that no significant improvements are achieved with this alternative setting.

The reconstruction employing only one type of sensor revealed that each of the wall quantities considered does not contribute in the same way to the reconstruction of each velocity component. Moreover, the implications due to the lack of availability of sensor resolution are addressed. In general, pressure measurements would provide better flow reconstructions than the wall shear stresses if data could be sampled at sufficiently high spatial resolution. However, this choice might bring an important penalization on the estimation of $u^+$ close to the wall, where $\tau_{w_x}$ performs remarkably better. Instead, when the resolution of the sensors is lower, $\tau_{w_x}$ is preferred. We address this to its wall patterns being significantly larger, thus a higher proportion of its energetic spectrum is preserved in the downsampled data.

Finally, the robustness of the 3D-GAN is assessed concerning the noise present in the sensing process at the wall, which would be present in a practical implementation of this methodology. As the amount of noise considered in the sensors increases, the accuracy of predictions is reduced. The reduction in accuracy is not the same for all the sensor arrangements considered. Finer resolutions are more affected by the noise introduced, while the 3D-GAN is more robust to noise under coarser wall inputs.

**Author contribution** All authors contributed to the study conception and design. Conceptualization: A.C., A.I., S.D.; Methodology: A.C., A.I., S.D.; Software: A.C.; Formal analysis: A.C.; Investigation: A.C.; Resources: A.I, S.D.; Data curation: A.C.; Writing—original draft preparation: A.C.; Writing - review and editing: A.C., A.I., S.D.; Visualization: A.C.; Supervision: A.I., S.D.; Project Administration: A.C., A.I., S.D.; Funding acquisition: A.C., S.D. All authors read and approved the final manuscript.

**Funding** Open Access funding provided thanks to the CRUE-CSIC agreement with Springer Nature. A.C. acknowledges funding from the Spanish Ministry of Universities under the Formación de Profesorado Universitario (FPU) 2020 program. The project received funding from the European Research Council (ERC) under the European Union's Horizon 2020 research and innovation program (grant agreement No 949085). Views and opinions expressed are however those of the authors only and do not necessarily reflect those of the European Union or the European Research Council. Neither the European Union nor the granting authority can be held responsible for them. Funding for APC: Universidad Carlos III de Madrid (Agreement CRUE-Madroño 2024).

**Data and code availability** Data related to this work and trained models of the neural networks are openly available at https://doi.org/10.5281/zenodo.13587745. Codes developed in this work are openly available at https://github.com/erc-nextflow/3D-GAN-limited-sensors.

**Declarations**

**Conflict of interest** The authors report no conflict of interest.

**Ethical approval** Not applicable.

**Consent for publication** Not applicable.